\documentstyle[12pt]{article}
\begin{document}

\baselineskip=20pt

\title{Necessary and Sufficient Conditions for Local Unitary Equivalence of Multi-qubit States}

\author{A. M. Martins\\
Departamento de F\'{i}sica, Instituto Superior T\'{e}cnico, \\
Universidade de Lisboa, 1049-001
Lisboa, Portugal }

\date{}

\maketitle

\begin{abstract}

We derive necessary and sufficient conditions for the LU-equivalence of two general (pure or mixed) $n$-qubit states as well as we determine the local unitary operators connecting them.  Almost all relevant information is contained in the $1$-qubit reduced matrices of the multiqubit states under investigation  Our technique relies on identifying {\it ab initio} all local symmetries and the corresponding local cyclic unitary operators. To derive the above conditions we use the reference forms of the multiqubit states whose definition requires the diagonalization of the 1-qubit reduced matrices. Based on those conditions we propose a straightforward protocol to decide wether or not two $n$-qubit states are LU-equivalent. \\\

PACS number(s) 03.67.-a, 03.67.Mn, 03.65.Aa, 02.20.Hj

\end{abstract}

\newpage  

\section{Introduction}

Measuring and classifying quantum entanglement  has been the object of extensive research work. The motivations are related to applications in quantum information and computation tasks \cite{Bennett1992, Bennett1993} as well as to the foundations of quantum physics \cite{Ariano2010, Pusey2012}. An exhaustive bibliography about these different aspects can be found in a recent review article by Horodecki and al. \cite{Horodecki2009}

A very fruitful approach to understand entanglement, was launched by the seminal work of Linden and al. \cite{Popescu1997, Popescu1998} who first used group-theoretic methods to classify entanglement in multi-qubit systems through their classes of local unitary (LU) equivalent states. Two quantum states that can be transformed into each other by LU operations, have the same amount of entanglement and are characterized by a complete set of polynomials, invariant under those local unitary transformations. Several authors analyzed the LU-equivalence along the paradigm of the polynomial invariants \cite{Grassl1998}-\cite{Jost2012}, however this approach becomes less and less operational as the number of subsystems increases. Kraus in \cite{Kraus2010} launched a new paradigm to the study of the LU-equivalence of multipartite pure states based on the concept of  standard form. More recently other techniques have been proposed to tackle the multipartite LU-eqivalence problem, like the matrix realignment and partial transposition \cite{Jost2013} and the generalized Bloch representation \cite{Jost2014}. 

In this work we derive necessary and sufficient conditions for the LU-equivalence of two general (pure or mixed) $n$-qubit states as well as we identify the local unitary operators connecting them. The core of our approach relies on the $1$-qubit reduced states of the given multiqubit states. We start by identifying the possible existence of local symmetries which are related with the invariance of the $1$-qubit reduced states under local unitary operators. Such local operators, named {\it cyclic} or {\it noneffective} \cite{Fu2006,Gharibian2008,Illuminati2011}, belong to the stabilizer subgroup of the reduced state that is left invariant under its action and may originate nonlocal effects in the global multiqubit state. Maximally mixed $1$-qubit reduced states are fixed by the whole group of local unitary operators, i.e., by $G= SU(2)^{\otimes  n}$. Expressing any non maximally mixed $1$-qubit state in the Pauli basis, allows us to identify the subgroup of unitary operators that leave the state invariant as being isomorphic to the subgroup of the $3$-dimensional rotations around the Bloch vector of the state. The next step consists in transforming each $1$-qubit reduced states to its diagonal form by the action of unitary operators, which hereinafter are used to compute the {\it reference forms} of the two multiqubit states. Finally, we derive the relation obeyed by the reference forms of the $n$-qubit states when they are LU-equivalent. Based on this relation we develop an operational way, in a form of a protocol, to decide the LU-equivalence of two $n$-qubit states. 

Our technique differs from the one proposed in the recent work \cite{Jost2014}. There the authors follow a different line of reasoning based on the singular value decomposition of the $n$-qubit states and the local symmetries are detected by the degeneracies of the corresponding eigenvalues. Moreover, the necessary and sufficient conditions that we have derived allow the derivation of explicit expressions for the local unitary operators that underlie the protocol proposed in the present work.

The paper is organized as follows. In Section 2, we use the generalized Pauli basis to express the multiqubit states and show how to identify local symmetries and the associated local cyclic operators. In Section 3, we define the reference form of a multiqubit state and derive necessary and sufficient conditions for LU-equivalence. The process here developed takes into account the different types of local symmetry and provides explicit expressions for the local unitary operators. In Section 4, we present an operational way, in the form of a protocol, to decide wether two given $n$-qubit states are LU-equivalent. In Section 5 we exemplify how to apply the technique here develloped to decide the LU-equivalence of a pair of pure as well as of a pair of mixed states. Finally we conclude in Section 6.

\section{Local symmetries}

A suitable choice of the basis set  to develop the density matrices may simplify considerably solving specific physical problems, or may help to identify new properties of the system. In this work, where systems are formed by $n$ similar $2$-level constituents, and where the partial trace operators play a determinant role, the natural choice of a basis set is the generalized Pauli vector basis. 

Let ${\cal V}_j$ denote the $4$-dimensional Hilbert space of $2 \times 2$ Hermitian matrices. A convenient basis for ${\cal V}_j$ is ${\cal  B}_j= \{\sigma_{\alpha_j} ; {\alpha_j}=0,1,2,3 \}$, where  $ \sigma_{\alpha_j}( \alpha_j=1,2,3)$ represents the usual Pauli matrices, and $\sigma_0 = {\bf 1}$, is the $2 \times 2$ identity matrix. Using in ${\cal V}_j$ the Hilbert-Schmidt inner product  $(\sigma_{\alpha_j }, \sigma_{\alpha_j^{\prime}}) =Tr\{ \sigma_{\alpha_j } \sigma_{\alpha_j^{\prime}}  \}  = 2 \delta_{\alpha_j , \alpha_j^{\prime}}  $, then ${\cal  B}_j $ is an orthogonal basis set. We are going to consider the set $ {\cal B}_{{\cal V}^{\otimes n}} = \{ \sigma_{\vec \alpha} \}$, where
\begin{equation}\label{vector2}
 \sigma_{\vec \alpha} = \otimes_{j=1}^{n}  \sigma_{\alpha_j}
\end{equation} 
The vector index ${\vec \alpha} =(\alpha_1, \alpha_2,... ,\alpha_n)$ is a $n$-tuple containing the $n$ indices $\alpha_j$. There exist $4^{n}$ such matrices all being traceless, except for $\sigma_{\vec 0 }= {\otimes}_{j=1}^n {\bf 1}_j$, which corresponds to the $2^n \times  2^n $ identity matrix with trace $Tr \{ \sigma_{\vec 0 } \} = 2^{n}$.

$ {\cal B}_{{\cal V}^{\otimes n}} $ is an orthogonal basis set of the complex $4^{n}$-dimensional Hilbert-Schmidt vector space ${\cal V}^{\otimes n} =\otimes_{j=1}^n {\cal V}_j $. Every complex square matrix, $(2^{n} \times 2^{n})$, can be seen as a vector $\bf v$, uniquely written in the form
\begin{equation}\label{vector}
{\bf v} =  \sum_{\vec \alpha} v_{\vec \alpha} \,\ \sigma_{\vec \alpha}
\end{equation}
where the components $v_{\vec \alpha} $ are given by 
\begin{equation}\label{components}
v_{\vec \alpha} =  \frac{1}{2^n} Tr\{ \sigma_{\vec \alpha} \,\ {\vec v}  \}
\end{equation}

Any $n$-qubit quantum state $\rho=\sum_{\vec \alpha} r_{\vec \alpha}  \sigma_{\vec \alpha} \in {{\cal V}^{\otimes n}} $, must be hermitian, $\rho = \rho^{\dag}$, positive definite, $\rho \geq 0$, and normalized $Tr \{ \rho \} =1$. These requirements on $\rho$ impose certain constraints to the components $ r_{ \vec \alpha} $: (a) $\forall_{\vec \alpha}, r_{\vec \alpha} \in \Re$,  (b) $r_{\vec 0} = \frac{1}{2^n} $, (c) $r_{\vec \alpha} = \frac{1}{2^n} Tr\{ \sigma_{\vec \alpha} \,\ \rho \}$ and (d) $\sum_{ \vec \alpha} r_{ \vec \alpha}^{2} \leq 1 $, the equality is attained for pure states.

The translated vector, ${\bar \rho }= \rho -  {\bf 1}^{ \otimes n} /2^{n}$,( ${\bf 1}^{ \otimes n} = \otimes_{j=1}^{n}  {\bf 1}_j$),  characterizes completely the quantum state $\rho$ and is the well known {\it generalized Bloch vector representation} of dimension $(4^{n} - 1)$.

The reduced density matrix of the $k$- qubits $({i_1 ,..,i_k})$ is given by
\begin{equation}\label{partial}
 \rho_{i_1 ,..,i_k}  =Tr_{n/  \{{i_1 ,..,i_k}\}} \{ \rho \}  \,\,\,\,\,\,\,\,\,\,\,\,\  ( i_l =1,...,n)
\end{equation}
where $Tr_{n/  \{{i_1 ,..,i_k}  \}} \{ . \}$ is the partial trace operator over $(n-k), (k < n)$ qubits, except qubits $({i_1 ,..,i_k})$. For instance, when $(k=1)$ eq.(\ref{partial}) gives the $1$-qubit density matrix $\rho_{i}$ and when $k=2$, it gives the $2$-qubit density matrix $\rho_{i,j}$. 

A local unitary transformation $U  \in G$ acts on a $n$-qubit state $\rho$ via the adjoint action,
\begin{equation}\label{LU}
\rho_U = ad \,\ U [ \rho] = U \rho U^{\dag} = \left( \otimes_{j=1}^{n} U_j \right) \rho  \left( \otimes_{j=0}^{n-1} U_{n-j}^{\dag} \right) 
\end{equation}
where $G= SU(2)^{\otimes  n}$ is a $3n$-dimensional Lie group and ${\cal L} = su(2) \oplus su(2) \oplus  ... \oplus su(2) $ is the corresponding Lie algebra. The set ${\cal B}_{\cal L} =  \{ {\bf b}_{\alpha_i}= \otimes_{k=1}^{i-1}  {\bf 1}_k  \otimes \sigma_{\alpha_i}  \otimes_{k'=i+1}^n {\bf 1}_{k'} ; \,\   \alpha_i = 1,2,3\} $ and $ i=1,..,n $, is a basis set for ${\cal L}$ whose elements are the generators of $G$. 

When two states $\rho_U $ and $\rho$ are LU-equivalent, then their $k$-qubit reduced states $ \rho_{U_{i_1 ,..,i_k}}$ and $\rho_{i_1 ,..,i_k}$ are related by
\begin{equation}\label{1equiv}
 \rho_{U_{i_1 ,..,i_k}}= \left( \otimes_{j=1}^{k} U_{i_j} \right) \rho_{i_1 ,..,i_k} \left( \otimes_{j=0}^{k-1} U_{i_{k-j}}^{\dag}  \right)
 \end{equation}
For a single qubit this expression reduces to 
 \begin{equation}\label{1equiv1}
 \rho_{U_{i}}=  U_{i}  \rho_{i }  U_{i}^{\dag}  
  \end{equation}
Two situations can happen:\\
1) $\forall_{i=1,...,n}$, $\rho_{U_{i}} \neq  \rho_{i }$, i.e., there is no local symmetry.  Note that when $\rho_i \neq \rho_{U_i} $, then $\rho_i \neq \frac {{\bf 1}}{2} $.  \\
2) There is at least one $l$ such that $ \rho_{U_{l}} = \rho_l $, i.e., qubit $l$ exhibits a local symmetry: \\
 2.a) {\it Weak}, when $\rho_l \neq \frac{{\bf 1}}{2}$,  \\
 2.b) {\it Strong}, when $\rho_l =\rho_l^{*} = \frac{{\bf 1}_l}{2}$ is the maximally mixed $1$-qubit state. 
 
When there is a local symmetry in qubit $i$, {\it weak} or {\it strong}, then . 
\begin{equation}\label{cyclic}
[\rho_{i} , U_{i}^{*}] = 0   \,\,\   \Leftrightarrow  \,\,\,\    \rho_i= U_{i}^{*} \rho_i U_{i}^{* \dag }
 \end{equation}
$U_{i}^{*}$ is a {\it cyclic} local unitary operator \cite{Fu2006,Gharibian2008}. The set of local unitary operators ${ U}_i^{*} \in SU(2)$  obeying the cyclic condition (\ref{cyclic}) is the stabilizer subgroup  of $\rho_i$, named $S(\rho_i)$.  When $\rho_i$ is maximally mixed, then the stabilizer subgroup is the entire group $G$. Using eq.(\ref{cyclic}) in eq.(\ref{1equiv1}) we obtain

\begin{equation}\label{1equiv10}
 \rho_{U_{i}}= ( U_{i} U_{i}^{*} ) \rho_i ( U_{i}^{* \dag }  U_{i}^{\dag} ) 
  \end{equation}
This is, any local operator $U_{i} U_{i}^{*} $, belonging to the left coset of the stabilizer subgroup of $\rho_i$, with respect to $U_i$, is also a LU operator connecting the states $\rho_i^{\prime} $ and $\rho_i$. When $U_i \in S(\rho_i)$ then $U_{i} U_{i}^{*}   \in S(\rho_i)$. The indeterminacy in the local unitary operator can, in principle, be solved due to global effects in multiqubit states. 
  
A generic local unitary operator $U_i \in SU(2)$ is a three real continuous parameter operator, $U( \varphi_i , \phi_i ,  \theta_i  )=  e^{i {\vec s}_i .{\vec \sigma} (i) }   = \cos ( \frac{\varphi_i}{2} ) {\bf 1}_i +i \sin ( \frac{\varphi_i}{2}) {\hat n}_{ {\vec s}_i} . {\vec \sigma} (i)$, with the following matrix form

\begin{equation}\label{uni}
U_i = U( \varphi_i , \phi_i ,  \theta_i  )=   
 \left(  \begin{array}{cc}
                             \cos \frac{\varphi_i}{2} + i \cos \theta_i  \sin \frac{\varphi_i }{2} &  \sin \frac{\varphi_i  }{2} \sin \theta_i  e^{- i \phi_i }    \\
                            - \sin \frac{\varphi_i }{2} \sin \theta_i  e^{ i \phi_i }             &  \cos \frac{\varphi_i }{2} - i \cos \theta_i  \sin \frac{\varphi_i }{2}
                             \end{array}
                           \right)
\end{equation}\\
${\hat n}_{ {\vec s}_i} = {\vec s}_i /  \parallel  {\vec s}_i   \parallel  =(n_{i_1} , n_{i_2}, n_{i_3})  \equiv ( \cos \phi_i  \sin \theta_i ,  \sin \phi_i  \sin \theta_i, \cos \theta_i )$ is a unit vector in the 3-dimensional Euclidian space (Bloch space of qubit $i$), parametrized by the azimuthal angle, $0 \leq \phi_i \leq 2 \pi $, and the polar angle, $0 \leq \theta_i \leq \pi $. The third parameter is $\varphi_i = 2 \parallel  {\vec s}_i  \parallel$ ($ 0 \leq \varphi_i  \leq \pi $). 

Any non maximally mixed $1$-qubit density matrix $\rho_i $ can be expressed in the Bloch basis  by
\begin{equation}\label{1qubitB}
\rho_i = \frac{1}{2} ({\bf 1}_i + {\vec r} (i)  \cdot  {\vec \sigma (i)} )
\end{equation}
where ${\vec r} (i) \neq 0 $ is the 3-dimensional Bloch vector, $ {\vec \sigma} (i)  = ( \sigma_{1_i} , \sigma_{2_i}, \sigma_{3_i} )$. In the Appendix 1 we show that the unitary operator $U_i^{*}$, that commutes with $\rho_i$ is given by $U_i^{*} = e^{i \omega_i  { \hat n_{{\vec r} (i)} } \cdot {\vec \sigma }(i)} $, with ${ \vec n_{{\vec r} (i)} }=  {\vec r} (i) / || {\vec r} (i) ||$ and  $\omega_i$ is a real continuous parameter such that $0 \leq \omega_i \leq 2 \pi $.  Invoking the local isomorphism between SU(2) and SO(3) we see that the unitary operator $U_i^{*}$, represents a rotation of an angle, $ \varphi_i=\frac{\omega_i }{2} $, around the vector ${\vec r} (i) $ of the Bloch sphere of qubit $i$. When ${\vec r} (i) = 0 $ then $\rho_i =  \frac{\bf 1}{2}$, and the unitary operator obeying condition (\ref{cyclic}), is a generic unitary operator (\ref{uni}), which represents a rotation of an angle $\varphi_i $ around an axis with direction ${\hat n}_{ {\vec s}_i}$ of the Bloch sphere.

\section{Necessary and sufficient conditions for LU-equivalence}

Let $V_i$ be the unitary operator that diagonalizes $\rho_i$, i.e., $\rho_i = V_i D_i V_i^{\dag}$, where $D_i = diag(\lambda_1 ,\lambda_2)$ is a diagonal matrix, ($\lambda_1 \leq \lambda_2$). Introducing $V_i D_i V_i^{\dag}$ in eq.(\ref{1equiv10}) we conclude that $ V_i^{\prime} =U_i U_i^{*} V_i $ is the unitary operator that diagonalizes $ \rho_{U_i} $, i.e. $\rho_{U_i} = V_i^{\prime} D_i V_i^{ \prime \dag}$ and $U_i =V_i^{\prime}  V_i^{  \dag} U_i^{* \dag} $. 

{\bf Proposition 1}: Let $\rho$ and $\rho^{\prime}$ be $n$-qubit states and let $D_i$ an $D_i^{\prime}$ be the diagonal matrices associated with the reduced states $\rho_i$ and $\rho_i^{\prime}$. If there is at least one $i$, such that $(D_i^{\prime} -D_i ) \neq 0$, then $\rho$ and $\rho^{\prime}$ are not LU-equivalent.

{\bf Proof:} If $\rho$ and $\rho^{'}$ are LU-equivalent then, $\forall_{i=1,..,n}$, there exist a local unitary operator $U_i$ such that $\rho_i^{\prime} =U_i \rho_i U_i$ which implies $D_i^{\prime} =D_i $. Therefore, if $( D_i^{\prime} -D_i ) \neq 0$ then the states $\rho^{\prime}$ and $\rho$ are not LU-equivalent. $\Box$

Let us define the {\it reference} forms $\rho^{(r)}$ and $ \rho^{\prime (r)}$ of two states $\rho$ and $\rho^{\prime}$ by
 \begin{equation}\label{standard2}
\rho^{(r)}=  (\otimes_{i=1}^{n} V_i^{ \dag} ) \,\   \rho \,\ (\otimes_{i=0}^{n-1} V_{n-i} )
 \end{equation}
 \begin{equation}\label{standard1}
 \rho^{\prime (r)} =  (\otimes_{i=1}^{n} V_i^{\prime \dag} ) \,\   \rho^{\prime} \,\  ( \otimes_{i=0}^{n-1} V_{n-i}^{\prime} )
 \end{equation}
whenever $\rho_i^{\prime} = \rho_i$ then $V_i^{\prime} =V_i$. If $ \rho_i = \rho_i^{*} =\frac {{\bf 1}}{2} $ then $V_i^{'} =V_i ={\bf 1}$. If $\forall_{i=1,..,n}$, $\rho_i$ is maximally mixed, then $\rho^{(r)}=  \rho $ and $\rho^{\prime (r)} =  \rho^{\prime} $.

Taking the partial trace over $(n-k)$ qubits, in eqs.(\ref{standard2}) and (\ref{standard1}), we obtain the reference forms of the reduced $k$-qubit state. They are   
 \begin{equation}\label{rstandard2}
\rho^{(r)}_{i_1 ,..,i_k}=  (\otimes_{j=1}^{k} V_{i_j}^{ \dag} ) \,\   \rho_{i_1 ,..,i_k} \,\ (\otimes_{j=0}^{k-1} V_{i_{k-j}})
 \end{equation}
 \begin{equation}\label{rstandard1}
\rho^{\prime (r)}_{i_1 ,..,i_k} =  (\otimes_{j=1}^{k}  V_{i_j}^{\prime \dag} ) \,\   \rho^{\prime}_{i_1 ,..,i_k} \,\  ( \otimes_{j=0}^{k-1}   V_{i_{k-j}}^{\prime} )
 \end{equation}
Based on the reference forms we are going to derive necessary and sufficient conditions for the LU-equivalence between two $n$-qubit states $\rho$ and $\rho^{\prime}$. 

{\bf Theorem 1:}  Let $\rho^{\prime}$ and $\rho$ be two $n$-qubit states.  The states $\rho^{\prime}$ and $\rho$ are LU-equivalent iff their reference forms are related by,
\begin{equation}\label{relation3}
\rho^{\prime (r)} = ( \otimes_{i=1}^{n}{\bar U}_i  ) \,\   \rho^{(r)}  \,\ ( \otimes_{i=0}^{n-1}{\bar U}_{n-i}^{\dag}  ) 
 \end{equation}
Where 
\begin{equation}\label{relation2}
  {\bar U}_i = U(\omega_i) = \left(  \begin{array}{cc}
                             e^{-i \omega_i} &  0 \\
                             0           &  e^{ i \omega_i} 
                             \end{array}
                           \right)
  \end{equation}
 if $\rho_i^{\prime}  \neq \frac{\bf 1}{2}$. And where $ {\bar U}_i = U(  \varphi_i , \phi_i  , \theta_i  )$, if $ \rho_i = \frac{\bf 1}{2}$. The parameter $\omega_i$ is computed in Corollary 3 and the parameters $(  \varphi_i , \phi_i  , \theta_i  )$ are computed in Corollary 4 of Appendix 2. When the states are LU-equivalent the local unitary operators relating them are given by $U_i = V_i^{\prime}{\bar U}_i V_i^{\dag}$, if $\rho_i^{\prime}  \neq \frac{\bf 1}{2}$ and by $U_i = U(  \varphi_i , \phi_i  , \theta_i  )$ if $\rho_i^{\prime} = \frac{\bf 1}{2}$.

{\bf Proof:}  {\it Sufficient condition}: If $\rho^{\prime}$ is LU-equivalent to $\rho$ then there is a $U=  \otimes_{i=1}^{n} U_i $ such that $\rho^{\prime}= \rho_U = U \rho U^{\dag}$. Using the definitions (\ref{standard2}) and (\ref{standard1}), we show that the reference forms of $\rho$ and of $\rho^{\prime} =\rho_U$, are  related by (\ref{relation3}) where ${ \bar U}_i = V_i^{\prime \dag} U_i V_i$. If $\rho_i^{\prime}  \neq \frac{\bf 1}{2} $ then $U_i = V_i^{\prime} V_i^{\dag} U_i^{* \dag} $, therefore ${ \bar U}_i =  V_i^{\dag} U_i^{* \dag} V_i$. Using Proposition 3 of Appendix 2 we show that ${\bar U}_i$ is given by (\ref{relation2}). Finally, if $ \rho_i = \frac{\bf 1}{2}$ then $V_i^{\prime }= V_i = {\bf 1}$ and ${\bar U}_i =U_i^{* \dag} =U(  \varphi_i , \phi_i  , \theta_i  )$ given by eq.(\ref{uni}) with the parameters computed in Corollary 4.

{\it Necessary condition}: Let us assume that the reference forms $\rho^{\prime (r)}$ and $ \rho^{(r)} $ are related by (\ref{relation3}). Replacing $\rho^{\prime}$ given by eq.(\ref{standard1}) in the left hand side of eq.(\ref{relation3}) and solving for $\rho^{\prime }$ we obtain
\begin{equation}\label{relation6}
\rho^{\prime } = ( \otimes_{i=1}^{n} V_i^{\prime}{\bar U}_i V_i^{\dag} ) \,\   \rho  \,\ ( \otimes_{i=0}^{n-1} V_{n-i} {\bar U}_{n-i}^{\dag} V_{n-i}^{\prime \dag} ) 
 \end{equation}
where, by hypothesis, ${\bar U}_i$ is given: by (\ref{relation2}) if $\rho_i^{\prime}  \neq \frac{\bf 1}{2}$ and by $ {\bar U}_i = U(  \varphi_i , \phi_i  , \theta_i  )$, if $ \rho_i = \frac{\bf 1}{2}$. Eq.(\ref{relation6}) shows that $\rho$ and $\rho^{\prime}$ are LU-equivalent and at the same time it gives the local unitary operators $U_i$ that relate the two states. They are $U_i = V_i^{\prime}{\bar U}_i V_i^{\dag}$ if $ \rho_i \neq  \frac{\bf 1}{2}$, and the angle $\omega_i$ is computed as in Corollary 3. If $ \rho_i = \frac{\bf 1}{2}$ then $V_i^{\prime} = V_i = {\bf 1}$ and $U_i = {\bar U}_i =  U(  \varphi_i , \phi_i  , \theta_i  ) $ with the angles $( \varphi_i , \phi_i  , \theta_i  )$ computed as in Corollary 4.
$\Box$
 \\

If all the coefficients of the $2$-qubit correlation terms are null, we compute the parameters of the LU operators using the $3$-qubit correlation terms, reasoning as in the Corollaries 3 and 4. If all the coefficients till the $k$-qubit correlation terms are null we use the ($k+1$)-qubit correlation terms. 

{\bf Corollary 1}: Let $\rho^{\prime}$ and $\rho$ be two LU-equivalent, $n$-qubit states. Then the reference forms of the $k$-qubit reduced states $\rho^{\prime (r)}_{i_1 ,..,i_k} $ and $ \rho^{(r)}_{i_1 ,..,i_k}$, are related by
\begin{equation}\label{reduced3}
\rho^{\prime (r)}_{i_1 ,..,i_k} = (  \otimes_{j= 1}^{k} {\bar U}_{i_j}) \,\   \rho^{(r)}_{i_1 ,..,i_k}   \,\ ( \otimes_{j= 0}^{k-1} {\bar U}_{i_{k-j}}^{\dag})  
 \end{equation}
 {\bf Proof:} Taking the partial trace of eq.(\ref{relation3}), over $(n-k)$ qubits, except quibts $i_1 ,.., i_k$, we obtain (\ref{reduced3}). $\Box$ 
 
This Corollary is going to be used in Appendix 2 to derive parameters $\omega_i$ and $( \varphi_i , \phi_i, \theta_i)$ using $k$-qubit reduced standard forms. When not all the coefficients of the second order correlation terms are null our technique is very easy to be applied to any multiqubit state. It becomes more complex when all the coefficients of the second order correlation functions are zero. In the next section we propose an operational way to check if two given multiqubit states are LU-equivalent and, in the affirmative case, it computes the local unitary operators relating them.

\section{Operational way to determine the LU-equivalence}

Let us assume that we were given two $n$-qubit states $\rho$ and $\rho^{\prime}$ and we were asked to check wether they are LU-equivalent. An operational way to answer to this question is given by the following protocol:

1 - Do $i=1$.

2 - Compute: (a) the reduced states $\rho_i$ and $\rho_i^{'}$, (b) the diagonal matrices $D_i= diag(\lambda_1, \lambda_2)$ and $D_i^{'}= diag(\lambda_1^{'}, \lambda_2^{'})$, with $\lambda_1 \leq \lambda_2$, $\lambda_1^{'} \leq \lambda_2^{'}$, (c) the corresponding unitary operators $V_i$ and $V_i^{'}$.

3 - Do i=i+1

4 - If $i=n+1$, go to step 5. If not, go to step 2.

5 - Do $i=1$

6 - Compute $(D_i^{'} - D_i) $. If it is null, go to step 7. If not, go to step 26. 

7 - Do $i=i+1$

8 - If $i=n+1$, go to step 9. If not, go to step 6.

9 - Do $i=1$

10 - Compute $(D_i^{'} - \frac{\bf 1}{2}) $. If it is null, go to step 11. If not, go to step 13. 

11 - Do $i=i+1$.

12 -  If $i=n+1$, go to step 22. If not, go to step 10.

13 - Do $j=i$.

14 - Do $i=1$.

15 - Compute $(\rho_i^{'} -  \frac{\bf 1}{2}) $. If it is null, go to step 19. If not, go to step 16.

16 - Compute $ \omega_i$ as in Corollary 3, compute ${\bar U}_i $ with eq.(\ref{relation2}) and $U_i = V_i^{\prime} {\bar U}_i  V_i^{\dag}$.

17 - Do $i=i+1$.

18 - If $i=n+1$ go to step 23 . If not go to step 15.

19 - Compute $( \varphi_i ,\phi_i , \theta_i )$ as in Corollary 4. Compute ${\bar U}_i =U ( \varphi_i ,\phi_i , \theta_i )$ and $U_i =U ( \varphi_i ,\phi_i , \theta_i )$.

20 - Do $i=i+1$.

21 - If $i=n$, go to step 23. If not, go to step 15.

22 - Compute $( \varphi_i ,\phi_i , \theta_i )$ by solving eqs.(\ref{ola14}). 

23 - Compute $\rho^{(r)}$ and $\rho^{\prime(r)}$ using eqs.(\ref{standard2}) and (\ref{standard1}).

24 - Compute $\left( \rho^{\prime(r)} - ( \otimes_{i=1}^{n}{\bar U}_i )\,\   \rho^{(r)}  \,\ ( \otimes_{i=1}^{n} {\bar U}_i^{\dag} )  \right)$. If null, go to step 25. If not, go to step 26.

25 - $\rho$ and $\rho^{'}$ are LU-equivalent. Write $U_i ; (i=1,..,n)$.

26 -$\rho$ and $\rho^{'}$ are not LU-equivalent.

\section{Explicit examples}

\subsection{Pure states}

We apply the above technique to the following pure states
\begin{equation}\label{example1}
| \psi \rangle =\frac{3}{5} |00 \rangle + \frac{4}{5} |11 \rangle
\end{equation}
\begin{equation}\label{example2}
| \psi^{\prime} \rangle =\frac{1}{10} |00 \rangle - \frac{7}{10} |01\rangle  - \frac{7}{10} |10\rangle + \frac{1
}{10} |11 \rangle
\end{equation}
The corresponding density operators expressed in the Pauli basis are
\begin{equation}\label{exampler1}
\rho =\frac{1}{4} \sigma_{0_1 , 0_2}   - \frac{7}{100} ( \sigma_{0_1 , {3}_2} +  \sigma_{{3}_1 ,0_2} )+ \frac{6}{25} ( \sigma_{1_1, {1}_2} -  \sigma_{{2}_1, 2_2}) + \frac{1}{4} \sigma_{3_1 , 3_2}
\end{equation}
\begin{equation}\label{exampler2}
\rho^{\prime} =\frac{1}{4} \sigma_{0_1 , 0_2} - \frac{7}{100} ( \sigma_{0_1 , 1_2} + \sigma_{1_1 , 0_2 } ) + \frac{6}{25}( \sigma_{{2}_1 , 2_2} -  \sigma_{3_1 , 3_2})+ \frac{1}{4} \sigma_{1_1 , {1}_2}
\end{equation}
and the 1-qubit density matrices are
\begin{equation}\label{example3}
\rho_1  =\frac{1}{2} {\bf 1}_{1}  - \frac{7}{50} \sigma_{{3}_1} \,\,\,\,\,; \,\,\,\ \rho_2  =\frac{1}{2} {\bf 1}_{2}  - \frac{7}{50} \sigma_{{3}_2}
\end{equation}
\begin{equation}\label{example4}
\rho_1^{\prime}  =\frac{1}{2} {\bf 1}_{1}  - \frac{7}{50} \sigma_{1_1} \,\,\,\,\  ; \,\,\,\    \rho_2^{\prime}  =\frac{1}{2} {\bf 1}_{2}  - \frac{7}{50} \sigma_{1_2}
\end{equation}
The diagonal matrices are $D_1 = D_2 =D_1^{\prime } =D_2^{\prime}= (\frac{9}{25},  \frac{16}{25} )$, the unitary operators that diagonalize $\rho_1$, $\rho_2$ are $V_1 =V_2= {\bf 1}$, and
\begin{equation}\label{example5}
V_1^{\prime} =V_2^{\prime}  = \frac{1}{ {\sqrt 2} }
\left(  \begin{array}{cc}
                             1 &  -1 \\
                            1  &  1  
                             \end{array}
                           \right)
\end{equation}
are the unitary operators that diagonalize $\rho_1^{\prime}$, $\rho_2^{\prime}$. The reference forms expressed in the Pauli basis are given by
\begin{equation}\label{example6}
\rho^{ (r)}  =\frac{1}{4} \sigma_{0_1 , 0_2}  - \frac{7}{100} ( \sigma_{0_1 , {3}_2} + \sigma_{{3}_1,0_2}) + \frac{6}{25}( \sigma_{1_1 , {1}_2} - \sigma_{{2}_1 , 2_2} )+ \frac{1}{4} \sigma_{3_1, 3_2}
\end{equation}
\begin{equation}\label{example7}
\rho^{\prime (r)}  =\frac{1}{4} \sigma_{0_1 ,0_2}  - \frac{7}{100} (  \sigma_{0_1 , {3}_2} + \sigma_{{3}_1 ,0_2}) - \frac{6}{25} ( \sigma_{1_1 , {1}_2} - \sigma_{{2}_1 , 2_2} )+ \frac{1}{4} \sigma_{3_1 , 3_2}
\end{equation}
In this example $r_{1_1 3_2  } =r_{2_1 3_2  } =0$ and we cannot use eqs.(\ref{omegak}), instead we use eqs.(\ref{omegak10}) which applied to these states are
\begin{equation}\label{example8}
\cos (2 \omega_1) \cos (2 \omega_2) - \sin (2 \omega_1) \sin (2 \omega_2) = -1 
\end{equation}
 \begin{equation}\label{example8}
\sin (2 \omega_1) \cos (2 \omega_2) + \cos (2 \omega_1) \sin (2 \omega_2) =0 
\end{equation}
or equivalently
 \begin{equation}\label{example9}
\cos \left[ 2 (\omega_1 + \omega_2) \right] = -1 \,\,\ ; \,\,\       \sin \left[ 2 (\omega_1 + \omega_2) \right]= 0
\end{equation}
with solution $ \omega_1 = \frac{ \pi }{2} - \omega_2$. All the parameters $\omega_1$ and $\omega_2$ verifying this relation give
 \begin{equation}\label{example10}
{\bar U}_1 \otimes {\bar U}_2 =
\left(  \begin{array}{cccc}
                             -i &  0 &  0   &0   \\
                            0 & i e^{-2 i \omega_1} & 0 & 0  \\
                            0  &   0  & - i e^{2 i \omega_1} & 0  \\
                            0  &   0  & 0  &  i
                             \end{array}
                           \right)
\end{equation}
and
 \begin{equation}\label{example11}
\rho^{\prime (r)}  - ( {\bar U}_1 \otimes {\bar U}_2  )  \rho^{(r)} ( {\bar U}_1^{\dag} \otimes {\bar U}_2^{\dag} )= 0
\end{equation}
The LU operators $U_1$ and $U_2$ such that $\rho^{\prime } =( U_1 \otimes U_2) \rho (U_1 \otimes U_2)^{\dag}$ are given by  
\begin{equation}\label{example12}
U_1 = V_1^{\prime} {\bar U}_1 V_1^{\dag}= \frac{1}{\sqrt 2}
\left(  \begin{array}{cc}
                             e^{-i \omega_1} &  -e^{i \omega_1}    \\
                           e^{-i \omega_1} &  e^{ i \omega_1}                        
                             \end{array}
                           \right)
\end{equation}
 \begin{equation}\label{example13}
U_2  =  V_2^{\prime} {\bar U}_2 V_2^{\dag} =  - i \frac{1}{\sqrt 2}
\left(  \begin{array}{cc}
                             e^{i \omega_1} &  e^{-i \omega_1}    \\
                           e^{i \omega_1} & - e^{ -i \omega_1}                        
                             \end{array}
                           \right)
\end{equation}

\subsection{Mixed states}

We apply the above technique to the following mixed states
\begin{equation}\label{examplem1}
\rho =\frac{1}{4} \sigma_{0_1 , 0_2}   - \frac{7}{150}( \sigma_{0_1 , {1}_2}  + \sigma_{1_1 , 0_2}) - \frac{7}{300} ( \sigma_{0_1 , 3_2} + \sigma_{3_1 , 0_2}) + 
 \frac{49}{3750}\sigma_{1_1 , {1}_2}  + \frac{49}{7500} \sigma_{3_1 , 3_2} 
\end{equation}
\begin{equation}\label{examplem2}
\rho^{\prime} =\frac{1}{4} \sigma_{0_1 , 0_2}  - \frac{7}{300} ( \sigma_{0_1 , {1}_2}  + \sigma_{3_1 , 0_2}  ) - \frac{7}{150} ( \sigma_{0_1 , 3_2} + \sigma_{1_1 , 0_2})  +  \frac{49}{7500} \sigma_{3_1 , {1}_2}+ \frac{49}{3750} \sigma_{1_1, 3_2}
\end{equation}
and the 1-qubit density matrices are
\begin{equation}\label{examplem3}
\rho_1  =\frac{1}{2} {\bf 1}_{1}- \frac{7}{75} \sigma_{1_1}- \frac{7}{150} \sigma_{3_1} \,\,\,\,\,; \,\,\,\ \rho_2  =\frac{1}{2} {\bf 1}_{2}- \frac{7}{75} \sigma_{1_2}- \frac{7}{150} \sigma_{3_2} 
\end{equation}
\begin{equation}\label{examplem4}
\rho_1^{\prime}  =\rho_1  \,\,\,\,\  ; \,\,\,\    \rho_2^{\prime}  =\frac{1}{2} {\bf 1}_{2}  - \frac{7}{150} \sigma_{1_2}- \frac{7}{75} \sigma_{3_2} 
\end{equation}
The diagonal matrices are $D_1 = D_2 =D_1^{\prime } =D_2^{\prime}= (\frac{1}{2}- \frac{7 \sqrt{5}}{150}  \,\ , \frac{1}{2}+ \frac{7 \sqrt{5}}{150} )$. Since the matrices of $\rho_1$, $\rho_1^{\prime}$ and $\rho_2$ coincide, then the unitary operator that diagonalizes them is 
\begin{equation}\label{examplemm5}
V_1= V_1^{\prime} =V_2  = 
\left(  \begin{array}{cc}
                             \frac{1+ \sqrt{5}}{\sqrt{10+2 \sqrt{5}}} &  \frac{1- \sqrt{5}}{\sqrt{10-2 \sqrt{5}}} \\
                           \frac{\sqrt{2}}{ \sqrt{5+ \sqrt{5}}}  &   \frac{\sqrt{2}}{ \sqrt{5- \sqrt{5}}}    
                             \end{array}
                           \right)
\end{equation}
and the unitary operator that diagonalizes $\rho_2^{\prime}$ is
\begin{equation}\label{examplem5}
V_2^{\prime} = 
\left(  \begin{array}{cc}
                             \frac{ 2 + \sqrt{5} }{ \sqrt{ 10 + 4 \sqrt {5}} } &   \frac{ 2 - \sqrt{5} }{ \sqrt{ 10 - 4 \sqrt {5}} }  \\
                            \frac{ 1 }{ \sqrt{ 10 + 4 \sqrt {5}} }  &   \frac{ 1 }{ \sqrt{ 10 - 4 \sqrt {5}} }  
                             \end{array}
                           \right)
\end{equation}

The reference forms expressed in the Pauli basis are given by
\begin{equation}\label{examplem6}
\rho^{ (r)}  = \frac{1}{4} \sigma_{0_1 , 0_2} - \frac{7 \sqrt{5}}{300}  ( \sigma_{0_1, {3}_2} + \sigma_{{3}_1, 0_2 } )
                 + \frac{49}{6250}\sigma_{1_1, 1_2} +  \frac{49}{18750}( \sigma_{1_1, 3_2} + \sigma_{3_1, 1_2} ) + \frac{147}{12500} \sigma_{3_1, 3_2}
\end{equation}
\begin{equation}\label{examplem7}
\rho^{ \prime (r)}  = \frac{1}{4} \sigma_{0_1 , 0_2}   - \frac{7 \sqrt{5}}{300} ( \sigma_{0_1 , 3_2} + \sigma_{{3}_1, 0_2 } )
                 - \frac{49}{6250}\sigma_{1_1 , 1_2} +  \frac{49}{18750} ( \sigma_{1_1, 3_2} - \sigma_{3_1, 1_2}) + \frac{147}{12500} \sigma_{3_1, 3_2}
\end{equation}
with $r_{1_1 3_2  } =r_{3_1 1_2  } = r_{1_1 3_2 }^{\prime} =  - r_{3_1 1_2  }^{\prime}= 49/ 18750 $, $ r_{2_1 3_2  }=  r_{3_1 2_2  } =r_{2_1 3_2 }^{\prime}=  r_{3_1 2_2 } ^{\prime} =0$. Introducing these coefficients in eqs. (\ref{omegak}) 
 we obtain
\begin{equation}\label{examplem8}
\cos (2 \omega_1)= 1 \,\,\ ; \,\,\   \sin (2 \omega_1) =0 \,\ ; \,\ \cos (2 \omega_2)= -1 \,\,\ ; \,\,\   \sin (2 \omega_2) =0
\end{equation}
therefore $\omega_1 =0$ and $ \omega_2 = \pi /2$. Introducing these angles in the operators ${\bar U}_i$ given by (\ref{relation2}) we obtain $ {\bar U}_1 = \bf{1}$ and ${\bar U}_2 = -i \sigma_3$, 
and
 \begin{equation}\label{examplem11}
\rho^{\prime (r)}  - ( {\bar U}_1 \otimes {\bar U}_2  )  \rho^{(r)} ( {\bar U}_1^{\dag} \otimes {\bar U}_2^{\dag} )= 0
\end{equation}
The LU operators $U_1$ and $U_2$ such that $\rho^{\prime } =( U_1 \otimes U_2) \rho (U_1 \otimes U_2)^{\dag}$ are given by $U_i=V_i^{\prime} {\bar U}_i V_i^{\dag} $, and we obtain
 \begin{equation}\label{examplem12}
U_1= {\bf 1} \,\,\,\   ;  \,\,\,\    U_2 =\frac{1}{\sqrt 2}( \sigma_{3_1} + \sigma_{1_1} ) 
\end{equation}

 \section{Concluding remarks}
 
 In this work we have derived necessary and sufficient conditions for the LU-equivalence of $n$-qubit states as well as explicit expressions for the local unitary operators that connect two LU-equivalent states. 
 
We have recognized and used the special role played by the $1$-qubit reduced density matrices in the detection of the LU-equivalence between $n$-qubit states, therefore the core of our technique relays on identifying, {\it ab initio}, all possible local symmetries or, equivalently, all local unitary cyclic operators. The other important role played by the $1$-qubit reduced states lays in determining the unitary matrices that take them to its diagonal form.
 These unitary matrices are then employed to compute the reference forms of the two multiqubit states which are the main ingredient to derive the necessary and sufficient condition for LU-equivalence. Based on the $1$-qubit diagonal matrix we also derive a simple criterium for LU-equivalence. Moreover, in the absence of local {\it strong} symmetries on all qubits, the computation of the local unitary operators is of an extreme simplicity enabling us to derive explicit expressions for the local unitary operators whenever the coefficients of the second order correlation terms are not all zero. The necessary and sufficient conditions derived in this work allow us to propose an easily implementable protocol to check for the existence of LU-equivalence.

 The technique here developed can be applied to multiqudit states $(d \geq 2)$ if explicit forms for local cyclic operators are known.

 \appendix
 
 \section{Appendix 1}
 
  {\bf Proposition 2}: The commutation relation $ [U_i , \rho_i ]=0 $, where the $1$-qubit reduced state is, $\rho_i =  \frac{1}{2} \left( {\bf 1}+ {\vec r} (i) . {\vec \sigma}(i) \right) $, with ${\vec r}_i \neq 0$, is verified iff 
\begin{equation}\label{uni5}
U_i =U (\xi_i ) =  e^{i \omega_i {\hat n}_{{\vec r} (i )}.{\vec \sigma} (i)  }= \cos ( \omega_i ) {\bf 1}_i + i \sin (\omega_i ) {\hat n}_{ {\vec r}(i)} . {\vec \sigma} (i)
\end{equation}  
is a single parameter unitary operator where $\omega_i =  \xi_i \parallel {\vec r} (i) \parallel $ where $\xi_i \in \Re$.
  
{\bf Proof}: The cyclic condition is equivalent to $U_i  \rho_i U_i^{\dag} = \rho_i \Leftrightarrow U_i [ {\vec r} (i) . {\vec \sigma}(i) ] U_i^{\dag } = {\vec r} (i) . {\vec \sigma}(i)  $. Replacing $U_i $, by $e^{i {\vec s} (i) .{\vec \sigma} (i) } =  \cos ( \parallel {\vec s} (i)  \parallel  ) {\bf 1} +i \sin (\parallel {\vec s} (i)  \parallel  ) {\hat n}_{ {\vec s} (i) } . {\vec \sigma} (i)$, in the last equality, we get
\begin{equation}\label{comutador2}
[ { {\vec s}(i)} . {\vec \sigma} (i) , {\vec r} (i) . {\vec \sigma}(i) ] = 0
\end{equation}  
Computing the above commutator, we obtain
\begin{equation}\label{comutador3}
[ {\vec s} . {\vec \sigma}  , {\vec r} . {\vec \sigma} ] = \sum_{k=1}^{3} \sum_{l=1}^{3} s_{jk} r_{jl} [ \sigma_k , \sigma_l  ] = 2 i \left(\sum_{k=1}^{3} \sum_{l=1}^{3} s_{k} r_{l} \epsilon_{klu} \right) Ê\sigma_u
\end{equation}    
where $\epsilon_{klu}$ is the Levi-Civita symbol. After some straightforward calculations we show that the commutator will be zero iff, ${\vec s} (i)= \xi_i  {\vec r} (i)  $, i.e., iff the vector ${\vec s} (i)$ is proportional to the vector ${\vec r} (i) $ with $\xi_i \in \Re$.

\section{Appendix 2}

 {\bf Proposition 3:} Let $V_i$ be the unitary operator that diagonalizes $\rho_i = \frac{1}{2} ({\bf 1}_i + {\vec r} (i)  \cdot  {\vec \sigma (i)} )$, ${\vec r} (i)  \neq 0$ and let $U_i^{*}$ be a local unitary operator that commutes with $\rho_i$. Then ${ \bar U}_i = V_i^{\dag} U_i^{* \dag} V_i$ is given by eq.(\ref{relation2}). \\
 {\bf Proof:} Eq.(\ref{1qubitB}) can be expressed in the equivalent form 
 \begin{equation}\label{form1}
 {\hat n}_{{\vec r} (i)} \cdot  {\vec \sigma (i)} ) = \frac{1}{|| {\vec r} (i)|| }( 2 \rho_i - {\bf 1}_i )
 \end{equation}
 where ${\hat n}_{{\vec r} (i)} =\frac {{ \vec r} (i)}{|| { \vec r} (i) ||}$ . If $V_i$ diagonalizes $\rho_i $ then
  \begin{equation}\label{form2}
 V_i^{\dag} \left( {\hat n}_{{\vec r} (i)} \cdot  {\vec \sigma (i)} \right) V_i = \frac{1}{ || {\vec r}_(i) || } ( 2 D_i - {\bf 1}_i ) = \frac{1}{ || {\vec r}_(i) || } 
 \left(  \begin{array}{cc}
                             2 \lambda_1 -1 &  0 \\
                             0           &  -( 2 \lambda_1 -1 )
                             \end{array}
                           \right)
 \end{equation}
From where we obtain
  \begin{equation}\label{form3}
Tr \{ \left[ V_i^{\dag} \left( {\hat n}_{{\vec r} (i)} \cdot  {\vec \sigma (i)} \right) V_i \right]^2 \}= \frac{2}{ || {\vec r}_(i) ||^2 } ( 1 - 2 \lambda_i)^2 =2
 \end{equation}
therefore $|| { \vec r} (i) || = 2 \lambda_1 -1$. Introducing this result in eq.(\ref{form2}) we get
 \begin{equation}\label{form25}
 V_i^{\dag} \left( {\hat n}_{{\vec r} (i)} \cdot  {\vec \sigma (i)} \right) V_i = 
  \left(  \begin{array}{cc}
                             1 &  0 \\
                             0           &  -1 
                     \end{array}
                           \right)
 \end{equation}
If $U_i$ commutes with $\rho_i$ then, by Proposition 2, $U_i =U_i^{*}= \cos \omega_i  {\bf 1} + i \sin \omega_i  ({\hat n}_{ {\vec r}(i)} . {\vec \sigma} (i)) $, and 
\begin{equation}\label{form26}
 {\bar U}_i =V_i^{\dag} U_i^{ * \dag }  V_i = V_i^{\dag}  \left( \cos  \omega_i  {\bf 1} - i \sin \omega_i ( {\hat n}_{ {\vec r}(i)} . {\vec \sigma} (i) \right)  V_i = 
  \left(  \begin{array}{cc}
                            e^{ -i \omega_i} &  0 \\
                             0           &   e^{ i \omega_i}
                             \end{array}
                           \right)=U(\omega_i)
 \end{equation}
equally we show that  $  V_i^{\dag} U_i^{* }  V_i =U(\omega_i)^{\dag}$
$\Box$ \\

 {\bf Corollary 2:} Let $V_i$ be the unitary operator that diagonalizes $\rho_i = \frac{1}{2} ({\bf 1}_i + {\vec r} (i)  \cdot  {\vec \sigma (i)} )$, ${\vec r} (i)  \neq 0$ and let ${\bar U}_i$ be a unitary operator given by eq.(\ref{relation2}). Then the unitary operator $ U_i = V_i {\bar U}_i  V_i^{\dag}$ commutes with $\rho_i$ and is given by  $U_i =e^{i \omega_i {\hat n}_{{\vec r} (i )}.{\vec \sigma} (i)  }$. \\
 
 {\bf Proof :} The operator ${\bar U}_i$ given by (\ref{relation2}) can be written in the form
 \begin{equation}\label{form10}
 {\bar U}_i =  \cos ( \omega_i ) {\bf 1} + i \sin (\omega_i ) 
  \left(  \begin{array}{cc}
                            1 &  0 \\
                             0           &   -1
                             \end{array}
                           \right)
 \end{equation}
and 
\begin{equation}\label{form11}
U_i = V_i  {\bar U}_i  V_i^{\dag} =  \cos ( \omega_i ) {\bf 1} + i \sin (\omega_i ) 
V_i   \left(  \begin{array}{cc}
                            1 &  0 \\
                             0           &   -1
                             \end{array}
                           \right) V_i^{\dag}
 \end{equation}
If $V_i$ is the unitary operator that diagonalizes $\rho_i = \frac{1}{2} ({\bf 1} + {\vec r} (i)  \cdot  {\vec \sigma (i)} )$, ${\vec r} (i)  \neq 0$ then, by (\ref{form25}), 
\begin{equation}\label{form12}
V_i   \left(  \begin{array}{cc}
                            1 &  0 \\
                             0           &   -1
                             \end{array}
                           \right) V_i^{\dag} = {\hat n}_{{\vec r} (i)} \cdot  {\vec \sigma (i)}
 \end{equation}
which, introduced in (\ref{form11}), gives $U_i =  \cos ( \omega_i ) {\bf 1} + i \sin (\omega_i ) \left( {\hat n}_{{\vec r} (i)} \cdot  {\vec \sigma (i)} \right)= e^{i \omega_i  {\hat n}_{{\vec r} (i)} \cdot  {\vec \sigma (i)} }$. Finally, by Proposition 2, $[U_i ,\rho_i ] =0$. $\Box$

 The action of ${\bar U}_i$, given by (\ref{form26}), in the three Pauli Matrices is the following
\begin{equation}\label{sigma1}
 {\bar U}_i \sigma_1 {\bar U}_i^{ \dag} = \cos (2 \omega_i ) \sigma_1 + \sin (2 \omega_i ) \sigma_2
  \end{equation}
\begin{equation}\label{sigma2}
 {\bar U}_i \sigma_2 {\bar U}_i^{ \dag} = - \sin (2 \omega_i ) \sigma_1 + \cos (2 \omega_i ) \sigma_2
  \end{equation}\begin{equation}\label{sigma3}
 {\bar U}_i \sigma_3 {\bar U}_i^{ \dag} =  \sigma_3
  \end{equation}
  
 {\bf Corollary 3:} Let $\rho$ and $\rho^{\prime}$ be two $n$-qubits states. Let us assume that $\rho_i \neq  \frac{{\bf 1}}{2}$ and that there is at least a $j \neq i $ such that $\rho_j \neq  \frac{{\bf 1}}{2}$. If $\rho^{\prime}$ is LU-equivalent to $\rho$, then the angle $\omega_i$ in eq.(\ref{relation2}) is given by 
\begin{equation}\label{omegak}
\cos (2  \omega_i ) =  \frac{r^{ (r)}_{1_i, 3_j } r^{\prime (r)}_{1_i, 3_j } + r^{ (r)}_{2_i, 3_j } r^{\prime (r)}_{2_i , 3_j}}{r^{ (r)^2}_{1_i, 3_j }  + r^{ (r)^2}_{2_i, 3_j }} 
\end{equation}
\begin{equation}\label{omegak2}
\sin (2  \omega_i ) = \frac{r^{ (r)}_{2_i, 3_j } r^{\prime (r)}_{1_i, 3_j } -r^{ (r)}_{1_i, 3_j } r^{\prime (r)}_{2_i, 3_j }}{r^{ (r)^2}_{1_i, 3_j }  + r^{ (r)^2}_{2_i, 3_j }}
\end{equation}  
if  $r^{ (r)}_{1_i, 3_j } \neq 0$.  or $r^{ (r)}_{2_i, 3_j } \neq 0$. If  $r^{ (r)}_{1_i, 3_j } = r^{ (r)}_{2_i, 3_j } =0$ then the angle $\omega_i$ is given by the solution of the following system of linear equations  
\begin{equation}\label{omegak10}
 M {\vec x} = {\vec r}^{\,\ \prime  (r)} 
\end{equation}  
 where ${\vec r}^{\,\ \prime  (r)} = (r_{1_i ,1_j}^{\prime (r)}, r_{2_i ,2_j}^{\prime (r)},r_{1_i ,2_j}^{\prime (r)},r_{2_i ,1_j}^{\prime (r)} )^T $ and ${\vec x} = (x_1, x_2,x_3,x_4 )^T$ is the vector of the unknown variables with $x_1 = \cos (2 \omega_i) \cos (2 \omega_j )$, $x_2 = \cos (2 \omega_i) \sin (2 \omega_j )$, $x_3 = \sin (2 \omega_i) \cos (2 \omega_j )$ and $x_4 = \sin (2 \omega_i) \sin (2 \omega_j )$. $M$ is the matrix of the coefficients given by
\begin{equation}\label{form12}
M  =  \left(  \begin{array}{cccc}
                            r_{1_i , 1_j}^{(r)} &   -r_{1_i , 2_j}^{(r)}  &   -r_{2_i , 1_j}^{(r)} &  r_{2_i , 2_j}^{(r)}  \\
                              r_{2_i , 2_j}^{(r)} &   r_{2_i , 1_j}^{(r)}  &   r_{1_i , 2_j}^{(r)} &  r_{1_i , 1_j}^{(r)}  \\
                                r_{1_i , 2_j}^{(r)} &   r_{1_i , 1_j}^{(r)}  &   -r_{2_i , 2_j}^{(r)} &  -r_{2_i , 1_j}^{(r)}  \\
                                  r_{2_i , 1_j}^{(r)} &   -r_{2_i , 2_j}^{(r)}  &   r_{1_i , 1_j}^{(r)} &  -r_{1_i , 21_j}^{(r)}  
                             \end{array}
                           \right) 
 \end{equation}
where $r^{ (r)}_{\alpha_i, \alpha_j } = r_{0_1..0_{i-1} \alpha_i 0_{i+1} ..0_{j-1} \alpha_j 0_{j+1}..0_n}^{(r)} $ are the coefficients of the reference form $\rho^{(r)}$ along the the Pauli vector 
$ \left( \otimes_{k=1}^{i-1} {\bf 1}_k \otimes \sigma_{\alpha_i} \otimes_{k=i+1}^{j-1} {\bf 1}_k \otimes \sigma_{\alpha_j} \otimes_{k=j+1}^{n} {\bf 1}_k \right) $ and  $r^{ \prime (r)}_{\alpha_i, \alpha_j } $ are the analogous coefficients for $\rho^{(r) \prime}$.

 {\bf Proof:} If $\rho^{\prime}$ is LU-equivalent to $\rho$, then by Corollary 1, the reference forms of $2$-qubit reduced states verify the equation
 \begin{equation}\label{reduced4}
\rho^{\prime(r)}_{i,j} - ( {\bar U}_i \otimes {\bar U}_j  ) \,\   \rho^{(r)}_{i,j}   \,\ (  {\bar U}_i \otimes {\bar U}_j)^{\dag}  =0
 \end{equation}
Expressing the $\rho^{\prime (r)}_{i,j}$ and $\rho^{(r)}_{i,j}$ in the Pauli basis, this equation is equivalent to the following one   
\begin{equation}\label{reducedr2}
\sum_{\alpha_i , \alpha_j =1}^{3}  \left( r^{\prime ( r)}_{\alpha_i , \alpha_j} \sigma_{\alpha_i}\otimes \sigma_{\alpha_j}  -  r^{(r)}_{\alpha_i , \alpha_j} ({\bar U}_i \sigma_{\alpha_i} {\bar U}_i^{\dag }  ) \otimes ({\bar U}_j \sigma_{\alpha_j }{\bar U}_j^{\dag}) \right) =0
\end{equation}
where $r^{(r)}_{\alpha_i , \alpha_j} =r_{0_1 .. 0_{i-1}  \alpha_i  0_{i+1}..0_{j-1}  \alpha_j  0_{j+1}..0_n}$. Let us assume that $\rho_i \neq \frac{{\bf 1}}{2}$ and that there is a $j \neq i$ such that $\rho_j \neq \frac{{\bf 1}}{2}$. Then $ {\bar U}_i \sigma_{\alpha_i} {\bar U}_i^{ \dag} $ is given by eqs.(\ref{sigma1}, \ref{sigma2}, \ref{sigma3}). Attending to the linear independence of the Pauli basis vectors, whenever eq.(\ref{reducedr2}) is verified, then 
 \begin{equation}\label{ola1}
r^{\prime (r)}_{1_i, 3_j } =  r^{ (r)}_{1_i, 3_j } \cos (2 \omega_i )  + r^{ (r)}_{2_i, 3_j } \sin ( 2 \omega_i)   
\end{equation}
 \begin{equation}\label{ola2}
r^{\prime (r)}_{2_i, 3_j } = r^{ (r)}_{2_i, 3_j } \cos (2 \omega_i )  - r^{ (r)}_{1_i, 3_j } \sin ( 2 \omega_i)
\end{equation}
 \begin{equation}\label{ola3}
r^{\prime (r)}_{3_i, 3_j } = r^{ (r)}_{3_i, 3_j } 
\end{equation}
and eq.(\ref{omegak10}) is also verified. When $r^{ (r)}_{1_i, 3_j } \neq 0$ or $r^{ (r)}_{2_i, 3_j } \neq 0$ we can solve the two eqs.(\ref{ola1}), (\ref{ola2}) in order to $\cos (2 \omega_i )$ and $\sin ( 2 \omega_i)$ obtaining eqs.(\ref{omegak}) and (\ref{omegak2}). When $r^{ (r)}_{1_i, 3_j } = r^{ (r)}_{2_i, 3_j } =0$, then we solve the system of eqs.(\ref{omegak10}) in order to ${\vec x}$, which allows us to compute $\cos (2 \omega_k)$ and $\sin (2 \omega_k)$ with $(k=i,j)$. $\Box$

If all the second order coefficients are null we have to use 3-qubit reduced reference forms, this is, we compute
 \begin{equation}\label{reduced44}
\rho^{\prime(r)}_{i_1,i_2, i_3} - (  \otimes_{i= i_1}^{i_3} {\bar U}_i  ) \,\   \rho^{(r)}_{i_1,i_2, i_3}   \,\ (   \otimes_{i= i_1}^{i_3} {\bar U}_j)^{\dag}  =0
 \end{equation}
and proceed as in Corollary 3. We express $\cos \omega_i$ and $\sin \omega_i$ in terms of the three order coefficients $r^{ (r)}_{\alpha_{i_1}, \alpha_{i_2}, \alpha_{i_3} } = r_{0_1..0_{i_1-1} \alpha_{i_1} 0_{i_1+1} ..0_{i_2-1} \alpha_{i_2} 0_{i_2 +1} ..0_{i_3-1}  \alpha_{i_3} 0_{i_3 +1}..0_n}^{(r)} $. In general, if all the coefficients of order $k < n$ are null we  apply a similar procedure to the coefficients of order $(k+1)$.  

 {\bf Corollary 4:} Let $\rho$ and $\rho^{\prime}$  be $n$-qubits states with one or more $\rho_i = \frac{{\bf 1}}{2}$. Let us assume that there is at least a $j \neq i $ such that $\rho_j \neq  \frac{{\bf 1}}{2}$. If $\rho^{\prime}$ is LU-equivalent to $\rho$, then the 3 parameters $(\varphi_i , \phi_i , \theta_i  )$ of $U(\varphi_i , \phi_i , \theta_i  )$ defined in eq.(\ref{uni}) are given by
 \begin{equation}\label{angle1}
\varphi_i  =  \frac{1}{2} \arccos \left( \frac{{\vec r}_{3_j }^{ \,\ \prime} \cdot { \vec r}_{3_j }}{\parallel  {\vec r}_{3_j }^{ \,\ \prime}  \parallel . \parallel  {\vec r}_{3_j }  \parallel }\right)
 \end{equation}
\begin{equation}\label{angle2}
\theta_i =  \frac{1}{2} \arccos (n_{3_i} )  
\end{equation}
\begin{equation}\label{angle3}
\phi_i  =  \frac{1}{2} \arctan \left( \frac{n_{2_i}}{n_{1_i}} \right) 
\end{equation}
where ${\vec r}_{3_j }=(r_{3_j, 1_i}^{(r)}, r_{3_j, 2_i}^{ (r)},r_{3_j, 3_i}^{(r)})$ and ${\vec r}_{3_j }^{ \,\ \prime}= (r_{3_j, 1_i}^{\,\ \prime (r)}, r_{3_j, 2_i}^{\prime (r)},r_{3_j, 3_i}^{\prime (r)})$ if ${\vec r}_{3_j } \neq 0$. The three parameters $n_{k_i}$ are defined by 
 \begin{equation}\label{direction}
n_{k_i} =  \frac{{\vec r}_{3_j }^{ \,\ \prime} \wedge { \vec r}_{3_j }}{\parallel  {\vec r}_{3_j }^{ \,\ \prime}  \parallel . \parallel  {\vec r}_{3_j }  \parallel } \cdot {\vec e}_k
 \end{equation}
 ${\vec e}_k , (k=1,2,3)$ are unit vectors of the $3$-dimensional reference frame.
 
 {\bf Proof:} f $\rho^{\prime}$ is LU-equivalent to $\rho$, then by Corollary 3, the reference forms of $2$-qubit states verify the eq.(\ref{reducedr2}). If $\rho_i = \frac{{\bf 1}}{2}$ for some $i$ then, $ {\bar U}_i = U(\varphi_i , \phi_i , \theta_i  ) $ is the most general local unitary matrix given by ({\ref{uni}). If there is at least one $j \neq i$ such that $\rho_j \neq \frac{\bf 1}{2}$ then ${\bar U}_j \sigma_{3_j }{\bar U}_j^{\dag} = \sigma_{3_j }$. Attending to the linear independence of the Pauli vectors, whenever eq.(\ref{reducedr2}) is verified, then 
 \begin{equation}\label{ola6}
 \sum_{\alpha_i =1}^{3}     \left( r^{\prime ( r)}_{ \alpha_i , 3_j } \sigma_{\alpha_i} -  r^{(r)}_{ \alpha_i ,3_j }  ({\bar U}_i \sigma_{\alpha_i } {\bar U}_i^{\dag}) \right) \otimes \sigma_{3_j}  =0
\end{equation}
invoking the local isomorphism between SU(2) and SO(3), then the second term inside the square brackets of the last equation can be written
 \begin{equation}\label{ola7}
  \sum_{\alpha_i =1}^{3}    r^{(r)}_{ \alpha_i ,3_j }  ({\bar U}_i \sigma_{\alpha_i }{\bar U}_i^{\dag})= \sum_{\alpha_i =1}^{3}  {\bar r}^{(r)}_{\alpha_i , 3_j }  \sigma_{\alpha_i }
\end{equation}
with
 \begin{equation}\label{ola8}
 {\bar r}^{(r)}_{ \alpha_i ,3_j } = \sum_{\alpha_i^{\prime} =1}^{3} R_{\alpha_i ,\alpha_i^{\prime}} r^{(r)}_{\alpha_i^{\prime} , 3_j  }
\end{equation}
where $R_{\alpha_i ,\alpha_i^{\prime}} $ is a $3$-dimensional orthogonal matrix, more precisely, it is the rotation matrix ${\bf R}({\hat n}_i,\varphi_i)$ of an angle $\varphi_i$ about a unit vector ${\hat n}_i=(n_{1_i} ,n_{2_i} ,n_{3_i})$, that transform the vector ${\vec r}_{3_j }=(r_{1_i ,3_j}^{(r)}, r_{2_i ,3_j}^{ (r)},r_{3_i ,3_j}^{(r)})$ into the vector $ {\vec { \bar r} }_{3_j }= ({\bar r}_{1_i ,3_j}^{ (r)}, {\bar r}_{ 2_i ,3_j}^{ (r)}, {\bar r}_{3_i ,3_j}^{ (r)}) $. Introducing the left side of eq.(\ref{ola7}) in eq.(\ref{ola6}) we obtain ${\vec r}_{3_j }^{ \,\ \prime} =  {\vec { \bar r} }_{3_j }$, with $\varphi_i$ and $n_{3_i}$ given by eqs.(\ref{angle1}) and (\ref{direction}). Equations (\ref{angle2}) and (\ref{angle3}) come from the definition of $n_{1_i} = \cos \phi_i \sin \theta_i$, $n_{2_i} = \sin \phi_i \sin \theta_i$ and $n_{3_i} = \cos \theta_i$. 
$\Box$

If there are $(n-1)$ reduced density matrices such that $\rho_j = \frac {{\bf 1}}{2} $ and sole $\rho_k \neq \frac {{\bf 1}}{2}  (k \neq j ) $ then we use Corollary 4 to compute the $(n-1)$ cyclic operators $U_j$. The operator ${\bar U}_k$ will be determined after knowing at least one $U_j$, using Corollary 3. 

If all $1$-qubit reduced density matrices are maximally mixed it is also possible to obtain the local unitary operators $U(\varphi_i , \phi_i , \theta_i  )$ using eq.(\ref{reducedr2}) and
reasoning as in Corollary 4. Replacing (\ref{ola7}) in (\ref{reducedr2}), we obtain
 \begin{equation}\label{ola10}
\sum_{\alpha_i =1}^{3}  \sigma_{\alpha_i} \otimes  \sum_{\alpha_j =1}^{3}  \left(  r^{\prime ( r)}_{\alpha_i ,\alpha_j }  -   {\bar r}^{(r)}_{ \alpha_i , \alpha_j }   \right) ({\bar U}_j \sigma_{\alpha_j }{\bar U}_j^{\dag}) =0
\end{equation}
invoking again the local isomorphism between SU(2) and SO(3), we have
 \begin{equation}\label{ola11}
 \sum_{\alpha_j =1}^{3}     \left(  r^{\prime ( r)}_{\alpha_i , \alpha_j}  -   {\bar r}^{(r)}_{\alpha_i ,\alpha_j }   \right)  ({\bar U}_j \sigma_{\alpha_j }{\bar U}_j^{\dag})= \sum_{\alpha_j =1}^{3}  {\overline {\left(  r^{\prime ( r)}_{\alpha_i , \alpha_j}  -   {\bar r}^{(r)}_{\alpha_i ,\alpha_j }   \right) }}  \sigma_{\alpha_j }
\end{equation}
where
 \begin{equation}\label{ola12}
 {\overline {\left(  r^{\prime ( r)}_{\alpha_i , \alpha_j}  -   {\bar r}^{(r)}_{\alpha_j , \alpha_i}   \right) }} = \sum_{\alpha_j^{\prime} =1}^{3} R_{\alpha_j ,\alpha_j^{\prime}}^{\prime} \left(  r^{\prime ( r)}_{\alpha_j , \alpha_j}  -   {\bar r}^{(r)}_{\alpha_j , \alpha_j}   \right) 
\end{equation}
where $R_{\alpha_j ,\alpha_j^{\prime}}^{\prime}$ is also a $3$-dimensional orthogonal matrix in the Bloch sphere of qubit $j$. Going back to eq.(\ref{ola10}) we get
\begin{equation}\label{reducedr4}
\sum_{\alpha_i =1}^{3}   \sum_{\alpha_j =1}^{3}  {\overline {\left(  r^{\prime ( r)}_{\alpha_i , \alpha_j}  -   {\bar r}^{(r)}_{\alpha_i ,\alpha_j }   \right) }} \sigma_{\alpha_i}  \otimes  \sigma_{\alpha_j } =0\end{equation}
From where we obtain nine equations 
 \begin{equation}\label{ola13}
{\overline {\left(  r^{\prime ( r)}_{\alpha_i , \alpha_j}  -   {\bar r}^{(r)}_{\alpha_i ,\alpha_j }   \right) }} =0
\end{equation}
or equivalently
 \begin{equation}\label{ola14}
 {\overline {\left(  r^{\prime ( r)}_{\alpha_i , \alpha_j}  -   {\bar r}^{(r)}_{\alpha_i ,\alpha_j }   \right) }} = \sum_{\alpha_j^{\prime} =1}^{3} R_{\alpha_j ,\alpha_j^{\prime}}^{\prime}  \left(  r^{\prime ( r)}_{\alpha_i ,\alpha_j^{\prime} }  -   \sum_{\alpha_i^{\prime} =1}^{3} R_{\alpha_i ,\alpha_i^{\prime}} r^{(r)}_{ \alpha_i^{\prime}, \alpha_j }  \right) =0
\end{equation}
where the unknowns are the six angles $(\varphi_j , \phi_j , \theta_j  )$ and $(\varphi_i , \phi_i , \theta_i  )$ of the rotation matrices $R_{\alpha_i ,\alpha_i^{\prime}}$ and $R_{\alpha_j ,\alpha_j^{\prime}}^{\prime}$. We don't present explicit solutions of eqs.(\ref{ola14}), they will be easier to solve given concrete values for $ r^{\prime ( r)}_{\alpha_i , \alpha_j} $ and $r^{(r)}_{ \alpha_i^{\prime}, \alpha_j } $. Once known the parameters $(\varphi_i , \phi_i , \theta_i  )$ for one of the maximally mixed qubits, then the parameters for any other qubit $k \neq i$ will be computed very easily using eqs.({\ref{angle1})-(\ref{direction}).  

Once again, if the coefficients of the 2-qubit reference forms are null we have to use the coefficients of the $ 3^{rd}$-order and so on. The procedure to compute the angles $(\varphi_i , \phi_i , \theta_i  )$ get more complex as higher order coefficients have to be used.

\end{document}